\documentclass[proof]{pasj00}
\draft

\begin{document}
\SetRunningHead{Author(s) in page-head}{Running Head}

\title{A New Galactic Extinction Map in High Ecliptic Latitudes}

\author{%
Tsunehito \textsc{Kohyama}\altaffilmark{1},
			Hiroshi \textsc{Shibai}\altaffilmark{1},
			Misato \textsc{Fukagawa}\altaffilmark{1},
			Takahiro \textsc{Sumi}\altaffilmark{1},
			
			and
			
			Yasunori \textsc{Hibi}\altaffilmark{2}
			}
\altaffiltext{1}{Department of Earth and School, Graduate School of Science, Osaka University, 1-1 Machikaneyama-cho, Toyonaka, Osaka 560-0043, Japan}
\email{kohyama@iral.ess.sci.osaka-u.ac.jp, shibai@iral.ess.sci.osaka-u.ac.jp, misato@iral.ess.sci.osaka-u.ac.jp, sumi@iral.ess.sci.osaka-u.ac.jp}
\altaffiltext{2}{Advanced Technology Center, National Astronomical Observatory of Japan, 2-21-1 Osawa, Mitaka, Tokyo 181-8588}
\email{yasunori.hibi@nao.ac.jp}
 

%


\maketitle

\begin{abstract}
In this study, we derived a galactic extinction map in high ecliptic latitudes for $\mid \beta \mid \ > 30^\circ$. The dust temperature distribution was derived from the intensities at 100 and 140 $\mu$m with a spatial resolution of 5$'$. The intensity at 140 $\mu$m was derived from the intensities at 60 and 100 $\mu$m of the {\it IRAS} data assuming two tight correlations between the intensities at 60, 100, and 140 $\mu$m of the {\it COBE}/DIRBE data. We found that these correlations can be separated into two correlations by the antenna temperature of the radio continuum at 41 GHz. 

Because the present study can trace the 5$'$-scale spatial variation in the dust temperature distribution, it has an advantage over the extinction map derived by Schlegel, Finkbeiner, and Davis, who used the DIRBE maps to derive dust temperature distribution with a spatial resolution of $1^\circ$. We estimated the accuracy of our method by comparing it with that of Schlegel, Finkbeiner, and Davis. The spatial resolution difference was found to be significant. The area in which the significant difference is confirmed occupies 28\% of the region for $\mid \beta \mid \ > 30^\circ$.

With respect to the estimation of extragalactic reddening, the present study has an advantage over the extinction map derived by Dobashi\ (2011), which was based on the 2MASS Point Source Catalog, because our extinction map is derived on the basis of far-infrared emission. Dobashi's extinction map exhibits a maximum value that is lower than that of our map in the galactic plane and a signal-to-noise ratio that is lower than that of our map in high galactic latitudes. This significant difference is confirmed in 81\% of the region for $\mid \beta \mid \ > 30^\circ$.

In the areas where the significant differences are confirmed, the extinction should be estimated using our method rather than the previous methods.

\end{abstract}

\section{Introduction}

Radiation from extragalactic objects is reduced by galactic interstellar dust, whose effect becomes appreciable in the ultraviolet (UV)--optical wavelength region. The foreground emission by the dust is significant in the cosmic background radiation at the infrared-radio wavelength (CIB; CMB). Therefore, the galactic dust component must be subtracted along each line of sight for the study of the cosmic background radiation.

For this purpose, Schlegel, Finkbeiner, \& Davis\ (1998; hereafter SFD) created an all-sky galactic extinction map by using the all-sky far-infrared maps of the Infrared Astronomy Satellite ({\it IRAS}) and the Diffuse Infrared Background Experiment (DIRBE) on the Cosmic Background Explorer ({\it COBE}). The spatial resolutions of the {\it IRAS} and DIRBE data are 5$'$ and $0^\circ.7$, respectively ({\it IRAS} Explanatory Supplement; DIRBE Explanatory Supplement). SFD derived the dust temperature from the DIRBE data at 100 and 240 $\mu$m because the far-infrared intensity at wavelengths shorter than 100 $\mu$m includes the emissions from very small dust grains that are transiently heated to high temperatures because of the absorption of high-energy photons (\cite{key-1}; Draine \& Anderson 1985; D\'{e}sert, Boulanger, \& Puget 1990). SFD derived an extinction map using {\it IRAS} at 100 $\mu$m intensity and dust temperature derived from the DIRBE data. Therefore, this map may not trace dust variations in the scale of a few minutes, as reported by Arce \& Goodman\ (1999). SFD calibrated the dust column density value into the color excess by using the observed reddening of elliptical galaxies cataloged in Faber et al.\ (1989).

Several extinction maps have been derived on the basis of star catalogs for dark cloud detection. Dobashi et al.\ (2005) derived a visual extinction map with a 6$'$ spatial resolution by using the Digitized Sky Survey (DSS). Dobashi\ (2011) derived an all-sky extinction map with a 1$'$ -- 12$'$ spatial resolution using the 2 Micron All Sky Survey Point Source Catalog (2MASS PSC). In addition, he derived the color excess of a star by using the "X percentile method," which is a recent extension of the Near Infrared Color Excess method (NICE, Lada et al.\ 1994; Rowels \& Froebrich\ 2009). Compared with the NICE method, the X percentile method is more robust against contamination by foreground stars because it statistically corrects for the effects of stars in the foreground. He has renewed the estimation of the background star colors using a star catalog simulated by Robin et al.\ (2003). These methods measure the dust only in front of the stars and are based on near-infrared radiation. Therefore, the precision of these methods can degrade in directions of high extinction. In addition, the signal-to-noise ratio (S/N) can decrease in directions of small extinction, such as that in high galactic latitudes. Therefore, these methods may be ineffective for the correction of galactic extinction in extragalaxies.

We created a galactic extinction map of the Cygnus region ($80^\circ \ < l \ < 90^\circ,\ -4^\circ \ < b \ < 8^\circ$) using a new method (Kohyama et al.\ 2010) that derived the dust temperature distribution with a 5$'$ spatial resolution. We obtained the intensity at 140 $\mu$m from the {\it IRAS} data at 60 and 100 $\mu$m, assuming two tight color--color correlations identified between DIRBE data at 60, 100, and 140 $\mu$m. The spatial resolution of our dust temperature map was 10 times higher than that of SFD, and our map could detect the galactic dust behind the star shown in the DSS catalog.

In this study, we apply our method to a region of high ecliptic latitudes for $\mid \beta \mid \ > 30^\circ$. In contrast to what was previously reported (Kohyama et al.\ 2010), a high galactic latitudes region is included here. The extinction map derived in this study is compared with those of SFD and Dobashi\ (2011). The difference between this study and SFD in the high galactic latitudes region is expected to be smaller than that in the galactic plane because the dust distribution in high galactic latitudes is more uniform than that in the galactic plane. The difference between this study and Dobashi\ (2011) is considered to be smaller than that between our previous study and Dobashi et al.\ (2005) because the near-infrared study in Dobashi\ (2011) allows the detection of a larger extinction using the color excess of more distant stars than that in the optical study by Dobashi et al.\ (2005).

The Far-Infrared Surveyor (FIS) on the {\it AKARI} satellite has a spatial resolution less than 1$'$ (Murakami et al.\ 2007; Kawada et al.\ 2007). FIS has observed the all-sky intensities at 65, 90, 140, and 160 $\mu$m, and its diffuse intensity maps are scheduled for publishing. The High Frequency Instrument (HFI) on the Planck satellite has observed the all-sky intensities at 350 $\mu$m to 3 mm (Planck HFI Core Team 2011a, b). Planck collaboration\ (2011) derived the dust temperature with a 4$'$ spatial resolution. A comparison of the independent results of the FIS, HFI, and this study is expected to improve the accuracy of dust temperature and extinction measurements.

The far-infrared data and our analysis are shown in Section 2, the color--color correlations are determined and separated in Section 3, the maps of dust temperature and extinction are derived in Section 4, the result of this study is compared with those of SFD and Dobashi\ (2011) in Section 5, and the conclusion is provided in Section 6.

\newpage

\section{Analysis and Data}
\subsection{Analysis}
In general, interstellar space in the galaxy is optically thin in the far-infrared wavelength region. The far-infrared intensity is written by Equation (1).
\begin{equation}
I_{\nu}({\lambda})=\tau_{100\ \mu m}\times\left(\frac{100\ \mu m}{\lambda}\right)^\beta \times B_{\nu}(\lambda, T_{d}),
\end{equation}
where $I_{\nu}(\lambda)$ is the intensity at wavelength $\lambda$, $\tau_{100\ \mu m}$ is the optical depth at 100 $\mu$m, $\beta$ is the spectral emissivity index, $B_{\nu}(\lambda, T_{d})$ is the Planck function, and $T_{d}$ is the dust temperature. We adopted $\beta = 2$ according to the previous studies (Draine \& Lee\ 1984; Hibi et al.\ 2006; SFD, Kohyama et al.\ 2010). In addition, we adopted a one-zone model such that the dust temperature is assumed to be constant along each line of sight, according to the previous studies (SFD; Kohyama et al.\ 2010).

Color-correction should be applied to the observed data with the photometric bands of DIRBE and {\it IRAS} because their published all-band intensities were derived by assuming that the source spectrum $\nu I_\nu(\lambda)$ is constant. The conversion from the observed (color-uncorrected) intensity to the color-corrected intensity is given by Equation (2) with a color-correction factor of $K_{\lambda}(\beta,T_{d})$:
\begin{equation}
I_{\nu, c}(\lambda)=\frac{I_{\nu}(\lambda)}{K_{\lambda}\left(\beta,T_{d}\right)},
\end{equation}
where $I_{\nu,c}(\lambda)$ represents the color-corrected intensity. The color-correction factor $K_{\lambda}(\beta,T_{d})$ is determined from the spectral emissivity index and dust temperature if the spectrum is given by Equation (1). The color-correction factor is unique for each instrument, and those of the DIRBE and {\it IRAS} intensities are referenced in the respective explanatory supplement. In this study, the color-corrections are applied for the intensities at 100 and 140 $\mu$m, but not for that at 60 $\mu$m, as reported in previous studies (Hibi et al.\ 2006; Kohyama et al.\ 2010). This is because the spectral energy distribution shape around 60 $\mu$m cannot be described by Equation (1). This correction affects the intensities by a few percentage points, and is therefore not crucial for our results.

\subsection{Data}
The far-infrared data are summarized in Table 1. All data are presented in the Hierarchical Equal Area isoLatitude Pixelization (HEALPix) projection (G\'{o}rski et al.\ 2005). We set the pixel scale in this study to be as large as 1$'$.72, which corresponds to Nside = 2048; Nside is the pixel scale parameter of the HEALPix projection. The HEALPix pixel scale has previously been defined discretely (G\'{o}rski et al.\ 2005, Table 1). The Improved Reprocessing of the {\it IRAS} Survey (IRIS) by Miville-Desch\^{e}nes \& Lagache\ (2005) was used as the {\it IRAS} image. The IRIS images are available in the HEALPix projection with a 1$'$.72 pixel scale. The Zodi Subtracted Mission Average data (ZSMA) was used as the DIRBE image, as reported in Kohyama et al.\ (2010). The ZSMA images are available in the Quadrilateralized Spherical Cube (QSC) projection, and the data are binned to 393,216 pixels with a 19$'$.5 pixel scale. The direct conversion from the QSC to HEALPix 1$'$.72 pixel size was found to be inappropriate because it produced blank pixels and forced interpolations during the conversion process. Therefore, we first attempted to convert the QSC data to the HEALPix projection with a 27$'$.5 pixel scale (Nside = 128). Conversion to the larger pixel size was not a disadvantage for this study because the ZSMA images need to be spatially smoothed to achieve sufficient S/N (see Section 3-1 for details). The images were smoothed by Gaussian convolution with a 30$'$ Full Width at Half Maximum (FWHM) to reduce the jagged pattern caused by the conversion. The 1$\sigma$ noise level of DIRBE140 is 2.4 MJysr$^{-1}$ in the QSC projection (DIRBE explanatory supplement), and has accordingly become 1.4 MJysr$^{-1}$ after the processes described above.

The interplanetary dust (IPD) emission in both IRIS and ZSMA images was first subtracted on the basis of the model of Kellsal et al.\ (1998). Then, the CIB offsets at the three wavelengths were subtracted in this study. The CIB intensities employed in this study (shown in Table 2) are lower than the upper limits indicated by Hauser et al.\ (1998).

Table 3 shows the notations for {\it IRAS} and DIRBE intensities at 60, 100, and 140 $\mu$m. The {\it IRAS} intensity at 140 $\mu$m was derived in this study. 

\section{Color--Color Correlations}
Hibi et al.\ (2006) analyzed the ZSMA data in the galactic plane ($\mid b \mid \ < 5^\circ$) and identified two color--color correlations at 60, 100, and 140 $\mu$m. Kohyama et al.\ (2010) analyzed the ZSMA data in the Cygnus region and identified two correlations that differed slightly from those identified by Hibi et al.\ (2006). In this study, we analyzed the ZSMA data and identified two correlations in the region for $\mid \beta \mid \ > 30^\circ$. The high galactic latitudes region was included in this analysis. The S/N of DIRBE140 is lower than those of DIRBE60 and DIRBE100. In addition, the S/N of DIRBE140 in high galactic latitudes is lower than that in the galactic plane. Therefore, the DIRBE maps were smoothed spatially by Gaussian convolution to improve the S/N of DIRBE140. Two correlations are found, and we separated the two correlations by using the Wilkinson Microwave Anisotropy Probe ({\it WMAP}; {\it WMAP} Seven-Year Explanatory Supplement) all-sky map at 41 GHz produced by Jarosik et al.\ (2011). 

\subsection{Smoothing of DIRBE maps and Identification of Color--Color Correlations}

As shown in Table 4, we changed the FWHM of the Gaussian depending on the S/N of DIRBE140 because the maps smoothed with a large beam have the disadvantage of throwing away small-scale information in the high-S/N regions. 

Figure 1 shows the S/N distribution of DIRBE140 before the processes described in Section 2-2. The FWHM of the Gaussian for the smoothing is defined to achieve an S/N of DIRBE140 over 10 (Table 4), with which S/N Hibi et al.\ (2006) found tight correlations in color--color diagrams.

Figure 2 shows the color--color diagrams in the region for $\mid \beta \mid \ > 30^\circ$. Two correlations with the different inclinations are seen in the color--color diagrams. The galactic and magellanic cloud data points shown in the figure are color--corrected. We define the regions of the LMC and SMC as those for which IRAS100 is over 10 MJysr$^{-1}$ and 3 MJysr$^{-1}$ in the directions of LMC and SMC, respectively. 

\subsection{Division of Color--Color Correlations}
Figure 3 shows two color--color diagrams with antenna temperatures both lower and higher than 0.7 mK at 41 GHz. The former exhibited a slower inclination than the latter. In a manner similar to that reported by Kohyama et al.\ (2010), we described the former and latter as slow and steep cases, respectively. The correlation coefficient of the slow and steep cases are $-0.6$ and $-0.8$, respectively. The two color--color diagrams appeared to show tight relations. Hibi et al.\ (2006) separated the two correlations by using the 10 GHz radio continuum emission map ($-4^\circ \ < l \ < 56^\circ, \mid b \mid \ < 1^\circ.5$) produced by Handa et al.\ (1987). Hibi et al.\ (2006) and Hirashita, Hibi, \& Shibai\ (2007) indicated that the inclination of the correlation in the color--color diagram changes depending on the intensity of the free--free emission on the line-of-sight. Hibi et al.\ (2006) divided all data points into the two groups with the border of the 0.5 K antenna temperature at 10 GHz. The correlation lower and higher than 0.5 K is called as the main and sub correlation, respectively (Hibi et al. 2006). The 10 GHz radio continuum emission map traces the free--free emission, but is available only within a small part of the sky. Therefore, we employed the 41 GHz radio continuum map for separating the data points into the two correlation groups because it also traces the free--free emission that is available for the whole sky. Kohyama et al.\ (2010) separated the two correlations such that DIRBE60/DIRBE100 is 0.28. A specific value of DIRBE60/DIRBE100 cannot separate the two correlations found in this study.

\subsection{Fitting}
Hibi et al.\ (2006) and Kohyama et al.\ (2010) fitted the color--color correlations as a power-law function represented by
\begin{equation}
\frac{DIRBE140}{DIRBE100} = a\left(\frac{DIRBE60}{DIRBE100}\right)^{b}.
\end{equation}
We also fit the color--color correlation in Equation (3) using the least squares method with parameters of coefficient and power in the two color--color diagrams (Figure 3); the results are summarized in Table 5. We tried two cases for the fitting: for case I, each data is given a weight according to its 1$\sigma$ photometric error, while for case II, no weight is applied. According to Kohyama et al.\ (2010), only the power-law index $b$ will be used, but the coefficient $a$ will not be used in the following discussion (Equation 4).

The fitting results are shown in the color--color diagrams in Figure 3. As seen in the right panel of Figure 3, the slopes of the power-law function for the steep case are similar among cases I and II and previous studies. However, the difference between cases I and II is significant for the slow cases, as shown in the left panel. Case I represents the correlation of the DIRBE data with a relatively high S/N (over 10 for DIRBE140). Case I is consistent with the main correlation (Hibi et al.\ 2006) and the slow case derived in Kohyama et al.\ (2010). The two latter cases were derived for the galactic plane. It is reasonable to expect that case I is consistent with the two latter cases because high S/N data concentrate to the galactic plane. On the other hand, case II weighs on the data with smaller S/N (but it is still large enough), which are mostly seen above the galactic plane. We add the difference between the results for cases I and II to the uncertainty of our method because it is difficult to accurately determine the spatial boundary of cases I and II from the present data. Case II may hold above the galactic plane, while case I may hold at the galactic plane.

\subsection{Uncertainty of the Interplanetary Dust Emission Subtraction}
The appropriate subtraction of the IPD emission is difficult to achieve because the IPD emission alone has not been observed. To avoid an uncertainty of the IPD emission subtraction, we did not analyze the region for $\mid \beta \mid \ < 30^\circ$, which shows the systematic stripe features appeared in DIRBE60 (and IRAS60) in the ecliptic plane (DIRBE60; Hauser et al.\ 1998; IRAS60; Miville-Desch\^{e}nes \& Lagache\ 2005). The stripe features were parallel to the ecliptic plane and did not appear in the region for $\mid \beta \mid \ > 30^\circ$. Kellsal et al.\ (1998) predicted the IPD emission with 1\% precision; however, there is significant uncertainty regarding the stripe feature. For example, the uncertainty is about 50\% at the ecliptic longitudes of 180$^\circ$. The average of DIRBE60 adjacent to the stripe is 0.6 MJysr$^{-1}$ and inside the stripe is 0.3 MJysr$^{-1}$.

\newpage

\section{Galactic Extinction Map}

We derived IRAS140, dust temperature, and extinction for cases I and II. The calculation method was the same as that reported in Kohyama et al.\ (2010), except for the power-law values of the color--color correlations. The power-law values were determined on the basis of the antenna temperature at 41 GHz of the line of sight. IRAS140 was derived by the following equation:
\begin{equation} 
\frac{IRAS140}{IRAS100} = \frac{DIRBE140}{DIRBE100}\left( \frac{IRAS60}{IRAS100} \div \frac{DIRBE60}{DIRBE100} \right)^{b}.
\end{equation}
The dust temperature was derived from intensities at 100 and 140 $\mu$m, and the optical depth at 100 $\mu$m was derived from the dust temperature and the intensity at 100 $\mu$m. The optical depth was converted to the extinction at the $V$ band (i.e., $A_{V}$) assuming the extinction law of Mathis\ (1990).

$I_{\nu, c}(140\ \mu m)$ of {\it IRAS} is the color-corrected intensity, which is derived as described below. $I_{\nu}(140\ \mu m)$ of {\it IRAS} is derived by the color-uncorrected {\it IRAS} and the color-corrected DIRBE data by Equation (4). From these $I_{\nu}(140\ \mu m)$ and $I_{\nu}(100\ \mu m)$ of {\it IRAS}, the dust temperature is calculated once. The dust temperature is used for the color-correction of $I_{\nu}(100\ \mu m)$ of {\it IRAS} to obtain $I_{\nu,c}(100\ \mu m)$ of {\it IRAS}. Finally, $I_{\nu, c}(140\ \mu m)$ of {\it IRAS} is derived from $I_{\nu, c}(100\ \mu m)$ and $I_{\nu}(60\ \mu m)$ of {\it IRAS} by Equation (4), and the dust temperature is derived by $I_{\nu, c}(100\ \mu m)$ and $I_{\nu, c}(140\ \mu m)$ of {\it IRAS}.

The notations for $A_{V}$ shown in this paper are summarized in Table 6. Figures 4 and 5 illustrate the maps of dust temperature and extinction, respectively.

We employ the results based on case I for the following reasons. The correlations in case I are believed to represent a common dust property described below. The color--color correlations of case I are similar to the correlations identified by Hibi et al.\ (2006) and those of extragalaxies. Hibi et al.\ (2006) found that the color--color correlation of the magellanic clouds were similar to the main correlation, which is the slow case of Hibi et al.\ (2006). We also found that the slow case I correlation is similar to the main correlation. Moreover, Hirashita \& Ichikawa\ (2009) found that the color--color correlations of the blue compact dwarf galaxies observed by {\it AKARI} are similar to the correlations identified by Hibi et al.\ (2006). 

The difference between the results for case I and the color--color correlations found by the previous studies are not added to the uncertainty of our method. It is because that the difference between the results for cases I and II is larger than that between the results for case I and Hibi et al.\ (2006) and that between the results for case I and Kohyama et al.\ (2010) in both slow and steep cases. As demonstrated in Section 5, the $A_{V}$ difference between SFD and the result for case I and that between Dobashi\ (2011) and the result for case I are significant even if the difference between the results for cases I and II is added to the uncertainty of our method. A significant difference is also expected even if the difference between the results for case I and Hibi et al.\ (2006) and that between the results for case I and Kohyama et al.\ (2010) are added to the uncertainty of our method.

\section{Results and Discussion}

As described above, we have derived the maps of dust temperature and extinction with a 5$'$ spatial resolution on the basis of far-infrared intensities. The derived extinction map based on case I was compared with the maps of SFD and Dobashi\ (2011).

\subsection{Comparison with SFD}
The difference in the spatial resolution of the dust temperature is expected to create dispersion in the difference between this study and SFD, but it should not cause a systematic difference. We estimated our accuracy assuming that SFD's map is correct in a 1$^\circ$ or larger spatial scale. In addition, we assumed that the difference in spatial resolution of the dust temperature causes dispersion only in the difference. Figure 6 shows the comparison between the $A_{V}$ values of case I and SFD, in which dispersion difference is evident. $A_{V}$ (SFD) is derived from the $E (B-V)$ of SFD assuming $R_{V} = 3.1$. In particular, a systematic difference appears for $A_{V}$ (SFD) $>$ 1 mag. The mean and median of the $A_{V}$ (case I)/$A_{V}$ (SFD) ratio for $A_{V}$ (SFD) $<$ 1 mag are 1.05 and 0.96, respectively. On the other hand, $A_{V}$ (case I) is systematically larger than $A_{V}$ (SFD) for $A_{V}$ (SFD) $>$ 1 mag. The mean and median of the $A_{V}$ (case I)/$A_{V}$ (SFD) ratio for $A_{V}$ (case I) $>$ 1 mag are 1.21 and 1.13, respectively. From the assumption described above, we estimated the accuracy of $A_{V}$ (case I) to be as much as 5\% and 21\% for $A_{V}$ (SFD) $<$ 1 mag and $A_{V}$ (SFD) $>$ 1 mag, respectively.

The reason why $A_{V}$ (case I) differs systematically from $A_{V}$ (SFD) for $A_{V}$ $>$ 1 mag is described below. The region for $A_{V}$ $>$ 1 mag corresponds to lower galactic latitudes regions ($\mid b \mid \ < 15^\circ$). In these regions, the dust temperature varies in a small spatial scale and can become quite low and high in limited portions of the sky. High-temperature regions with scales of a few minutes cannot be resolved in the DIRBE maps. In contrast to the DIRBE maps, the {\it IRAS} maps resolve these high-temperature regions. Therefore, the {\it IRAS} dust temperature map shows a larger area of the high-temperature region when compared to the DIRBE dust temperature map. In these regions, the scatter diagram between the dust temperature distributions derived by the DIRBE maps and that derived by the {\it IRAS} maps will show that the dust temperature derived by the {\it IRAS} maps is systematically lower than that derived by the DIRBE maps. When the dust temperature decreases, the dust extinction increases. Therefore, the extinction derived by the {\it IRAS} maps becomes systematically larger than that derived by the DIRBE maps. SFD derived the dust temperature using the DIRBE maps, but we derived the dust temperature using the {\it IRAS} maps. Therefore, in these regions (i.e., $A_{V}$ $>$ 1 mag), the dust extinction derived by our method is systematically larger than that derived by SFD.

The difference in the spatial resolution of the dust temperature contributes to the horizontal dispersion in Figure 6; this trend changes at $A_{V}$ (SFD) $=$ 0.1 mag. The horizontal dispersion for $A_{V}$ (SFD) $<$ 0.1 mag appears relatively wider than that for $A_{V}$ (SFD) $>$ 0.1 mag. The wider dispersion may be caused by SFD's assumption that dust temperature is uniform in the small extinction region (i.e., in high galactic latitudes), and they therefore combined the DIRBE maps with background averages obtained at high galactic latitudes for $\mid b \mid \ > 75^\circ$. No such assumption was made in our method.

Figure 7 shows a histogram of $A_{V}$ (SFD) vs. $A_{V}$ (case I). The mean, median, and standard deviation are $-0.28$, $-0.001$, and 2.29 mag, respectively. Figure 8 shows the $A_{V}$ difference distribution of $A_{V}$ (SFD) vs. $A_{V}$ (case I). The absolute $A_{V}$ difference between $A_{V}$ (case I) and $A_{V}$ (SFD) in high galactic latitudes is smaller than that in the galactic plane because of the smaller extinction in high galactic latitudes. In addition, the dust distribution in high galactic latitudes is more uniform than that in the galactic plane, and the standard deviation in high galactic latitudes is significantly smaller than that in the galactic plane. The standard deviation is 0.10 mag for $\mid b \mid \ > 45^\circ$ and 2.66 mag for $\mid b \mid \ < 45^\circ$.

Figure 9 shows the uncertainty of our method estimated by the following equation:
\begin{equation}
\Delta A_{V} = \Delta(noise) + \Delta (case\ I\ -\ case\ II) + \Delta (accuracy\ of\ the\ case\ I).
\end{equation}
The first term is the $A_{V}$ 1$\sigma$ noise error calculated from the IRIS 1$\sigma$ noise at 60 and 100 $\mu$m. Photometric noise could cause the horizontal dispersion in Figure 6 in addition to the difference in the spatial resolution of the dust temperature. The second term is the difference between $A_{V}$ (case I) and $A_{V}$ (case II) calculated as $\mid A_{V}$ (case I) $-$ $A_{V}$ (case II)$\mid$. The third term is the accuracy of $A_{V}$ (case I) as estimated by the systematic difference between this study and SFD. The values of the third term are 0.05$A_{V}$ (SFD) and 0.21$A_{V}$ (SFD), for $A_{V}$ (SFD) $<$ 1 and $A_{V}$ (SFD) $>$ 1 mag, respectively. This is a dominant term for $A_{V}$ (SFD) $>$ 1 mag ($\mid b \mid < 15^\circ$). The second term is dominant for $A_{V}$ (SFD) $<$ 1 mag ($\mid b \mid > 15^\circ$). Near the galactic pole, the first term becomes equivalent to the second term because of the low S/N of the {\it IRAS} data. The extinction value should be estimated by our method rather than by SFD's method when the difference from $A_{V}$ (SFD) is larger than $\Delta A_{V}$ given in Equation (5). Figure 10 shows the area in which the $A_{V}$ difference between this study and SFD is larger than the $A_{V}$ uncertainty of our method, as shown in the following equation: 
\begin{equation}
\mid A_{V} (SFD) - A_{V} (case\ I)\mid\ >\ \Delta A_{V}.
\end{equation}
The area represented in Figure 10 occupies 28\% of the region for $\mid \beta \mid \ > 30^\circ$, and exists in not only the galactic plane but also high galactic latitudes. The area in which the significant difference is confirmed increases if we consider just case I or II because the nominal uncertainty in each case is smaller than the difference between $A_{V}$ (case I) and $A_{V}$ (case II). If we assume case I only, the area occupies 49\% of the region for $\mid \beta \mid \ > 30^\circ$.

\subsection{Comparison with Dobashi\ (2011)}
$A_{V}$ (2MASS) is converted from $E (J-H)$ derived by Dobashi\ (2011), assuming the extinction law of Cardelli, Clayton, \& Mathis\ (1989) as reported in Dobashi\ (2011). $E (J-H)$ is the most sensitive quantity in the color excesses derived by Dobashi (2011). The case of $E (J-H)^{50}_{Xm}$, used for the dark cloud survey in Dobashi\ (2011), is compared in this study (see the definition of $E (J-H)^{50}_{Xm}$ described in Dobashi 2011). The LMC and SMC were masked because Dobashi\ (2011) did not derive the extinction of the LMC and SMC.

Figure 11 shows the result of the comparison between this study and Dobashi\ (2011). The $A_{V}$ (2MASS) are saturated at $A_{V}$ (2MASS) $>$ 20 mag, indicating that the method of Dobashi\ (2011) has an upper limit of $A_{V}$ (2MASS) $=$ 20 mag. It is apparent this method is not useful for $A_{V}$ $>$ 20 mag even though the upper limit of $A_{V}$ $=$ 20 mag is four times as large as that of the visual extinction derived by Dobashi et al.\ (2005). The visual extinction derived by Dobashi et al.\ (2005) was saturated at $A_{V}$ $>$ 5 mag in the Cygnus region (Kohyama et al.\ 2010). $A_{V}$ (2MASS) is systematically smaller than $A_{V}$ (case I) for 1 -- 10 mag because galactic dust may exist behind the stars cataloged in the 2MASS PSC. Dispersion that appears in the vertical direction for the $A_{V}$ (case I) $<$ 1 mag is probably caused by the low S/N of $A_{V}$ (2MASS) in high galactic latitudes. In high latitudes, Dobashi\ (2011) employed a beam size that was larger than that in the galactic plane to achieve a constant noise level. The extinction in high latitudes was smaller than that in the galactic plane. The 1$\sigma$ noise level is 0.4 mag in $A_{V}$ (2MASS). The mean uncertainty of our method was 0.07 mag for $\mid b \mid \ > 45^\circ$.

Figure 12 shows the histogram of $A_{V}$ (2MASS) $-$ $A_{V}$ (case I). The mean, median, and standard deviation of the $A_{V}$ difference are $-0.58$, $-0.09$, and 3.69 mag, respectively. Figure 13 shows the distributions of the $A_{V}$ difference between this study and Dobashi (2011). As expected in the systematic difference in Figure 11, a systematic difference is seen in the galactic plane. The mean and median for $\mid b \mid \ < 45^\circ$ are $-0.75$ and $-0.10$ mag, respectively. Though galactic dust may exist behind the stars cataloged in the 2MASS PSC, $A_{V}$ (2MASS) is systematically larger than $A_{V}$ (case I) toward the galactic bulge. It indicates that Dobashi\ (2011) may overestimate the extinction toward the galactic bulge. A scatter appears in high galactic latitudes. The standard deviation for $\mid b \mid \ > 45^\circ$ is 0.34 mag.

Figure 14 shows the area in which the difference in $A_{V}$ obtained in this study and Dobashi\ (2011) is larger than the $A_{V}$ uncertainty of our method, as shown in the following equation:
\begin{equation}
\mid A_{V} (2MASS) - A_{V} (case\ I)\mid\ > \Delta A_{V}.
\end{equation}
The area represented in Figure 14 occupies 81\% of the region for $\mid \beta \mid \ > 30^\circ$. The $A_{V}$ (2MASS) and $A_{V}$ (case I) are in agreement at middle-galactic latitudes, which indicates that the extinction derived at near-infrared wavelengths above the middle-galactic latitudes may trace most of the dust column in our galaxy along a line of sight.

In the area represented in Figure 14, the extinction should be estimated using our method instead of Dobashi\ (2011)'s method with respect to the estimation of the extragalactic reddening. However, our method may not be suitable as an estimator of the reddening for objects located in the galaxy because the extinction derived in this study reflects the entire integration through the galaxy. Note that the reddening derived by Dobashi\ (2011) is larger than that derived by our method toward the galactic bulge as shown in Figure 13.

The area in which the significant difference is confirmed increases if we consider just case I or case II. If we assume case I only, the area occupies 90\% of the region for $\mid \beta \mid \ > 30^\circ$.

The $A_{V}$ comparison results are summarized in Table 7.

Future prospects are described below. Observation results of {\it AKARI}/FIS should be compared with the results of this study. FIS has observed the all-sky intensities at 65, 90, 140, and 160 $\mu$m with a spatial resolution less than 1$'$ (Murakami et al.\ 2007; Kawada et al.\ 2007). The FIS results are independent of the results of this study. The FIS observed data at 65 and 100 $\mu$m will be calibrated by the {\it IRAS} and DIRBE data. The FIS observation data at 140 $\mu$m can be calibrated by the DIRBE data. Therefore, with respect to the spatial resolution, the FIS calibration accuracy at 140 $\mu$m is 1$^\circ$. The accuracy is expected to increase upon comparison with this study. {\it Planck}/HFI showed the dust emission distribution with a 4$'$ spatial resolution. The spatial resolution was nearly equal to that obtained in this study. Planck collaboration (2011) derived the dust temperature distribution on the basis of intensities at 100 and 350 $\mu$m. The {\it IRAS} data were used as the intensities at 100 $\mu$m in the Planck collaboration (2011). Planck collaboration\ (2011) assumed a spectral emissivity index of $\beta = 1.8$. If a significant difference between {\it Planck}'s result and this study is confirmed, the validity of the spectral emissivity index of the dust emission should be examined. We will produce maps of the dust temperature assuming a spectral emissivity index of $\beta = 1.8$.

\section{Conclusion}
We derived a new galactic extinction map in high ecliptic latitudes for $\mid \beta \mid \ > 30^\circ$ on the basis of the far-infrared intensity and dust temperature with a spatial resolution of 5$'$. The dust temperature was derived on the basis of intensities at 100 and 140 $\mu$m. The intensity at 140 $\mu$m was derived from the intensities of the {\it IRAS} data at 60 and 100 $\mu$m, assuming color--color correlations between the intensities of the DIRBE data at 60, 100, and 140 $\mu$m.

We identified two tight correlations in the region for $\mid \beta \mid \ > 30^\circ$. The DIRBE maps were spatially smoothed to increase the S/N of DIRBE140, particularly in high galactic latitudes. We found that the two correlations can be separated by the antenna temperature of the radio continuum at 41 GHz.

Our method had a clear advantage over SFD because the spatial resolution of the dust temperature was 10 times higher than that of SFD. The accuracy was estimated to be as much as 5\% and 21\%, for $A_{V}$ (SFD) $<$ 1 and $A_{V}$ (SFD) $>$ 1 mag, respectively. A significant difference in $A_{V}$ was confirmed in the comparison between this study and SFD. The area in which the deviation was larger than the $A_{V}$ uncertainty of our method occupied 28\% of the region for $\mid \beta \mid \ > 30^\circ$. In this area, the extinction should be estimated using our method instead of SFD's method.

The extinction map derived in this study was compared with that derived by Dobashi\ (2011). The $A_{V}$ (2MASS) was saturated at $A_{V}$ $>$ 20 mag. A low S/N of $A_{V}$ (2MASS) appeared in high galactic latitudes. The S/N of $A_{V}$ (2MASS) was significantly smaller than that determined by this study in high galactic latitudes. The area in which the significant difference was confirmed occupied 81\% of the region for $\mid \beta \mid \ > 30^\circ$. In this area, with respect to the estimation of extragalactic reddening, the extinction should be represented by our method as opposed to Dobashi\ (2011)'s method.

\bigskip

All analyses described in this paper were carried out in Interactive Data Language (IDL) with the HEALPix package (K. M. G\' {o}rski et al.\ 2005, ApJ, 622, p759). The {\it COBE}/DIRBE and {\it WMAP} Seven-Year data were obtained from the Legacy Archive for Microwave Background Data Analysis (LAMBDA). The IRIS data were obtained from the IRIS project, and the SFD data were obtained from the Dust Maps project. The 2MASS extinction data were obtained from the Astro Laboratory of Tokyo Gakugei University. This study was supported by Grant-in-Aid for challenging Exploratory Research (22654023).



\newpage

\begin{table}[htb]
\begin{center}
\caption{Summary of the far-infrared data.}\label{tab:tab_1}
\begin{tabular}{llllllll}
\hline
\hline
 & \shortstack{Wavelength \\ \ }& \shortstack{Spatial \\ resolution}&\shortstack{1$\sigma$ \\ noise level } &\shortstack{Calibration \\ error} & \shortstack{IPD emission \\ model} & \shortstack{Projection method \\ used in this study} & \shortstack{Pixel scale \\ used in this study} \\
 & [$\mu$m] & [arcmin] & [MJysr$^{-1}$] & [\%] & & & [arcmin] \\
\hline
{\it IRAS} & 60 & 4$'$.0 & 0.03 & 10.5 & & & \\
{\it IRAS} & 100 & 4$'$.3 & 0.06 & 13.5 & & & \\
DIRBE & 60 & & 0.09 & 10.5 & \shortstack{Kellsal et al. \\ (1998)} & HEALPix & 1$'$.72\\ 
DIRBE & 100 & 42$'$ & 0.1 & 13.5 & & \\ 
DIRBE & 140 & & 2.4 & 10.6 & & \\ 
\hline
\multicolumn{1}{@{}l}{\rlap{\parbox[t]{.95\textwidth}{\small
{\it IRAS} data are the IRIS data (Miville-Desch\^{e}nes, \& Lagache 2005). The DIRBE data are ZSMA (DIRBE explanatory supplement). }}}
\end{tabular}
\end{center}
\end{table}

\begin{table}[htb]
\begin{center}
\caption{CIB intensities at 60, 100, and 140 $\mu$m.}\label{tab:tab_2}
\begin{tabular}{lll}
\hline
\hline
Wavelength [$\mu$m] & Intensity [MJysr$^{-1}$] & Reference \\
\hline
60 & 0.27 $\pm$ 0.05 & Miville-Desch\^{e}nes, Lagache, \& Puget\ (2002)\\
100 & 0.73 $\pm$ 0.03 & Planck collaboration\ (2011) \\
140 & 0.94 $\pm$ 0.16 & Matsuura et al.\ (2011) \\
\hline
\end{tabular}
\end{center}
\end{table}

\begin{table}[htb]
\begin{center}
\caption{Notations for the {\it IRAS} and DIRBE data.}\label{tab:tab_3}
\begin{tabular}{lll}
\hline
\hline
IRAS60 & intensity observed by {\it IRAS} at 60 $\mu$m \\
IRAS100 & same as above, but at 100 $\mu$m \\
IRAS140 & intensity derivied in this study at 140 $\mu$m\\
DIRBE60 & intensity observed by DIRBE at 60 $\mu$m.\\
DIRBE100 & same as above, but at 100 $\mu$m.\\
DIRBE140 & same as above, but at 140 $\mu$m.\\
\hline
\end{tabular}
\end{center}
\end{table}

\begin{table}[htb]
\begin{center}
\caption{Summary of the spatial smoothing process by Gaussian convolution used to achieve the S/N of DIRBE140 over 10.}\label{tab:tab_4}
\begin{tabular}{cccc}
\hline
\hline
\shortstack{S/N ratio \\ before the smoothing} & \shortstack{FWHM of a \\ Gaussian [deg]} &\shortstack{Spatial resolution \\ after the smoothing [deg]} & \shortstack{Expected S/N \\ ratio improvement}\\
\hline
$>$ 10 & - & 0.9 & $\times$ 1\\
5 -- 10 & 1.5 & 1.8 & $\times$ 2\\
2.5 -- 5 & 3.4 & 3.6 & $\times$ 4\\
1.3 -- 2.5 & 7.0 & 7.2 & $\times$ 8\\
$<$ 1.3 & 13.8 & 14.4 & $\times$ 16\\
\hline
\end{tabular}
\end{center}
\end{table}

\newpage

\begin{figure}
  \begin{center}
    \FigureFile(160mm,90mm){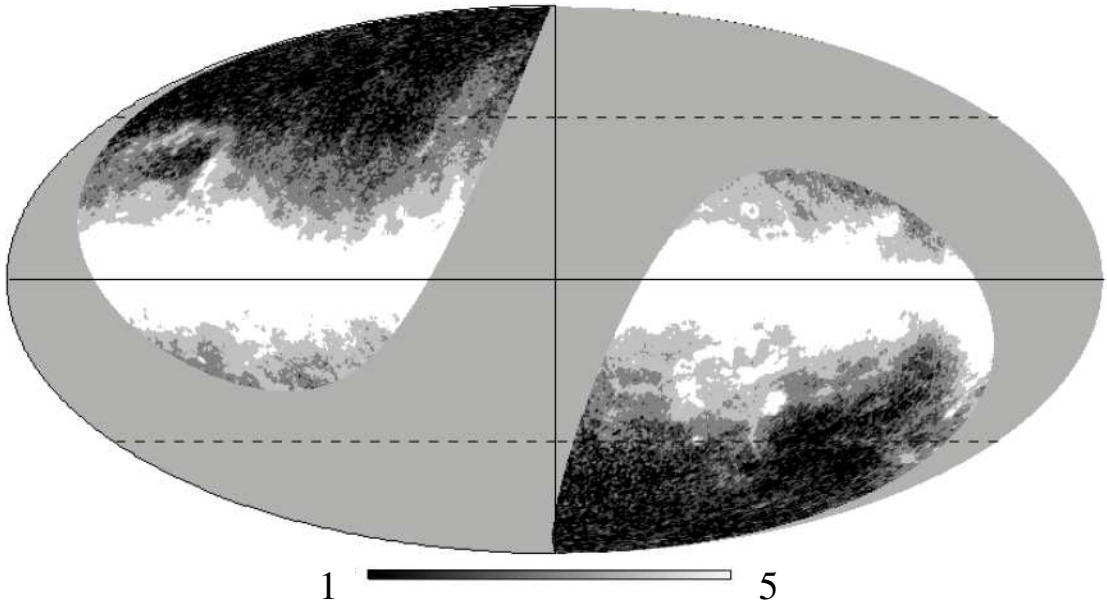}
  \end{center}
  \caption{Map of the S/N of DIRBE140, which was divided into five classes. The white region represents classes with S/N ratios higher than those of the black region, and the gray zone represents the region that is for $\mid \beta \mid \ < 30^\circ$. The distribution is shown in the Mollweide projection with the galactic coordinate. The dotted lines indicate the galactic latitudes of $\mid b \mid \ = 45^\circ$}\label{fig:fig_1}
\end{figure}

\newpage

\begin{figure}
  \begin{center}
    \FigureFile(160mm,80mm,clip){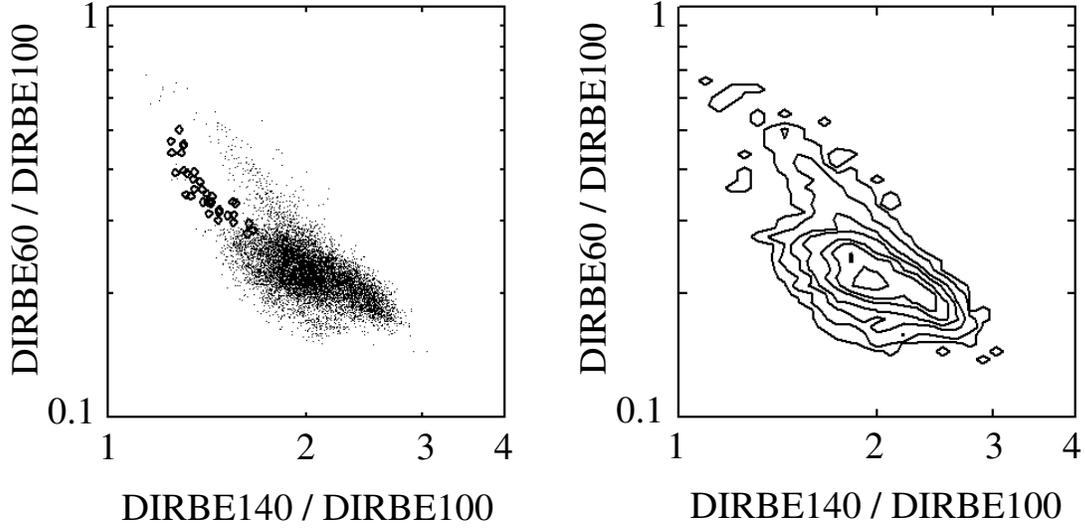}
  \end{center}
  \caption{(left) Color--color diagram of the region for $\mid \beta \mid \ > 30^\circ$. The DIRBE maps were smoothed to achieve an S/N of DIRBE140 over 10. The dots and squares represent galactic and magellanic data points, respectively. (right) The contour map illustrates the density of the data points on the diagrams. The bin size is set to 0.02 in logarithmic units. Contour lines are drawn according to the number of data points, including 0.5 (to express the boxes that have only one data point), and are 0.1N, 0.5N, N, 2N, 3N, and 5N on the diagram, where N = 40 indicates the average number of data points contained in each bin on this contour map. }\label{fig:fig_2}
\end{figure}

\begin{figure}
  \begin{center}
    \FigureFile(160mm,80mm,clip){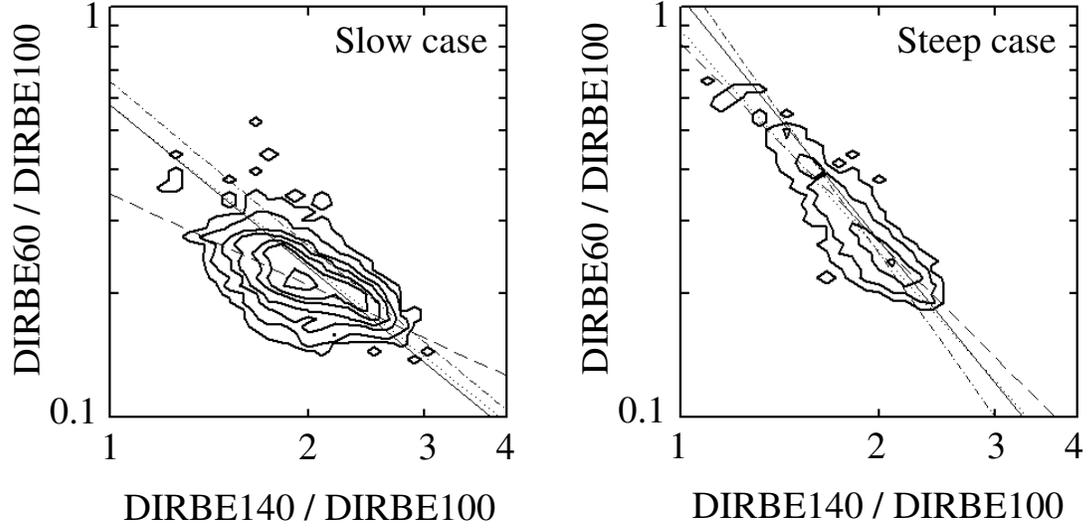}
  \end{center}
  \caption{Two color--color diagrams of the region for $\mid \beta \mid \ > 30^\circ$. (left) 41 GHz antenna temperature below 0.7 mK; (right) 41 GHz antenna temperature over 0.7 mK. Solid and dashed lines show the color--color correlations for cases I and II, respectively; dotted lines show the result in the galactic plane (Hibi et al.\ 2006), with the main correlation in the left and the sub correlation in the right; and dotted-dashed lines show the result in the Cygnus region (Kohyama et al.\ 2010). The 41 GHz radio continuum emission map was not used by Hibi et al.\ (2006) or Kohyama et al.\ (2010) to separate the two correlations they identified. The contour map includes the same parameters as those described in Figure 2.}\label{fig:fig_3}
\end{figure}

\clearpage

\begin{table}[htb]
\begin{center}
\caption{Fitting results for the color--color correlations obtained by a power-law function (Equation 3).}\label{tab:tab_5}
\begin{tabular}{lll}
\hline
\hline
Case & Slow case & Steep case \\
\hline
case I & $a= 0.659 \pm 0.003$ & $a= 1.03 \pm 0.02$ \\
weighted least squares fitting & $b = -0.767 \pm 0.003$ & $b = -0.508 \pm 0.004$ \\
($\mid \beta \mid \ > 30^\circ$) \\
\hline
case II & $a= 0.24 \pm 0.02$ & $a= 0.88 \pm 0.07$ \\
un-weighted least squares fitting & $b = -1.36 \pm 0.02$ & $b = -0.627 \pm 0.001$ \\
($\mid \beta \mid \ > 30^\circ$) \\
\hline
\hline
Hibi et al.\ (2006) & $a= 0.65$ & $a = 0.93$\\
Galactic plane  & $b = -0.78$ & $b = -0.56$\\
($\mid b \mid \ < 5^\circ$) & (main correlation) & (sub correlation)\\
\hline
Kohyama et al.\ (2010) & $a= 0.73$ & $a = 1.09$\\
Cygnus region & $b = -0.75$ & $b = -0.44$\\
($80^\circ < l < 90^\circ,\ -4^\circ < b < 8^\circ$)\\
\hline
\multicolumn{1}{@{}l}{\rlap{\parbox[t]{.95\textwidth}{\small
Errors in this study are shown in 1$\sigma$.}}}
\end{tabular}
\end{center}
\end{table}

\begin{table}[htb]
\begin{center}
\caption{Calculation methods for $A_{V}$.}\label{tab:tab_6}
\begin{tabular}{ll}
\hline
\hline
Notation & Method \\
\hline
$A_{V}$ (case I) & our method, assuming case I. \\
$A_{V}$ (case II) & our method, assuming case II. \\
$A_{V}$ (SFD) & Schlegel, Finkbeiner, \& Davis (1998)\\
$A_{V}$ (2MASS) & Dobashi (2011) \\
\hline
\end{tabular}
\end{center}
\end{table}

\begin{figure}
  \begin{center}
    \FigureFile(160mm,clip){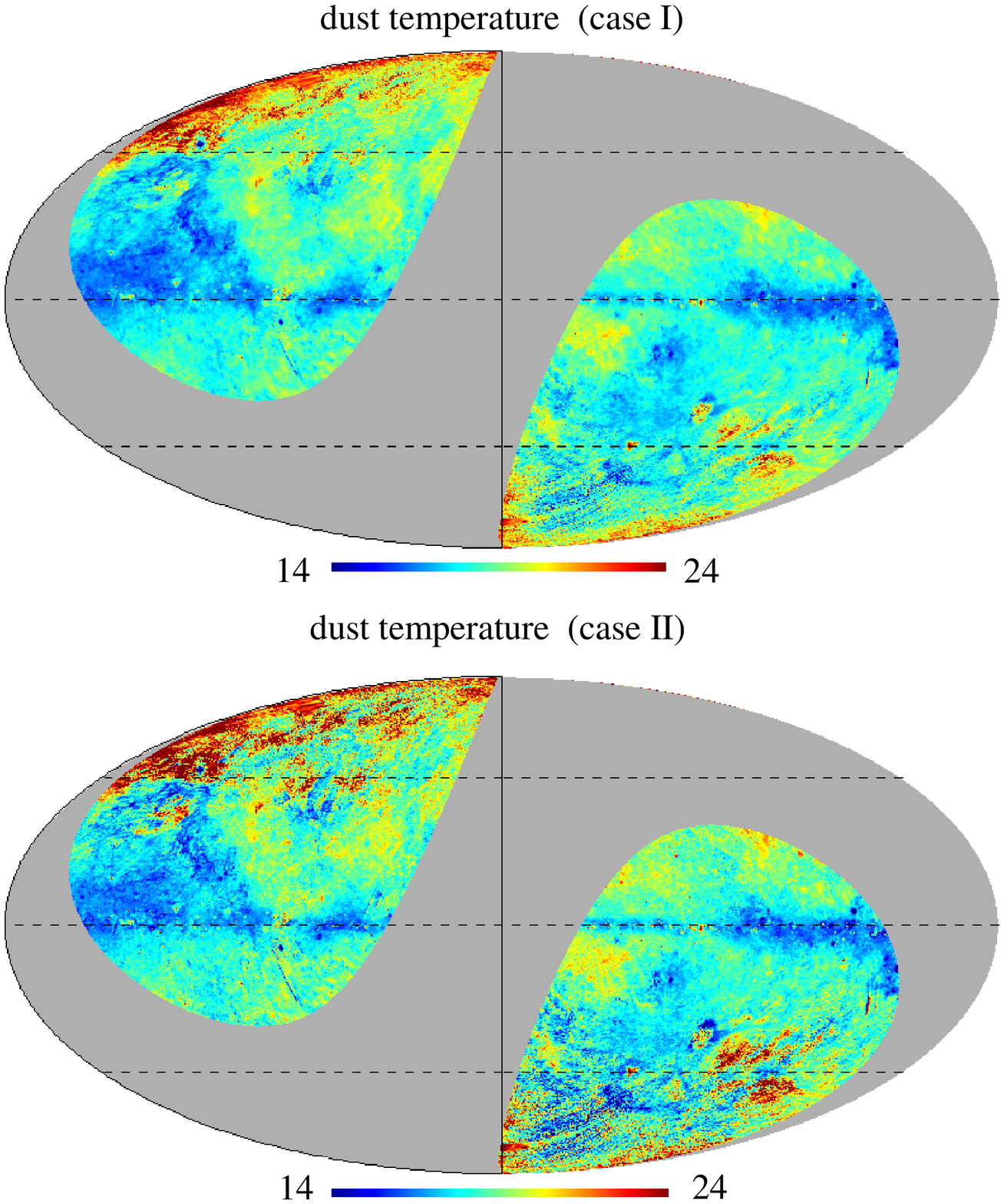}
  \end{center}
  \caption{Dust temperature distributions [K]. Top and bottom columns show results for cases I and II, respectively.}\label{fig:fig_4}
\end{figure}

\begin{figure}
  \begin{center}
    \FigureFile(160mm,clip){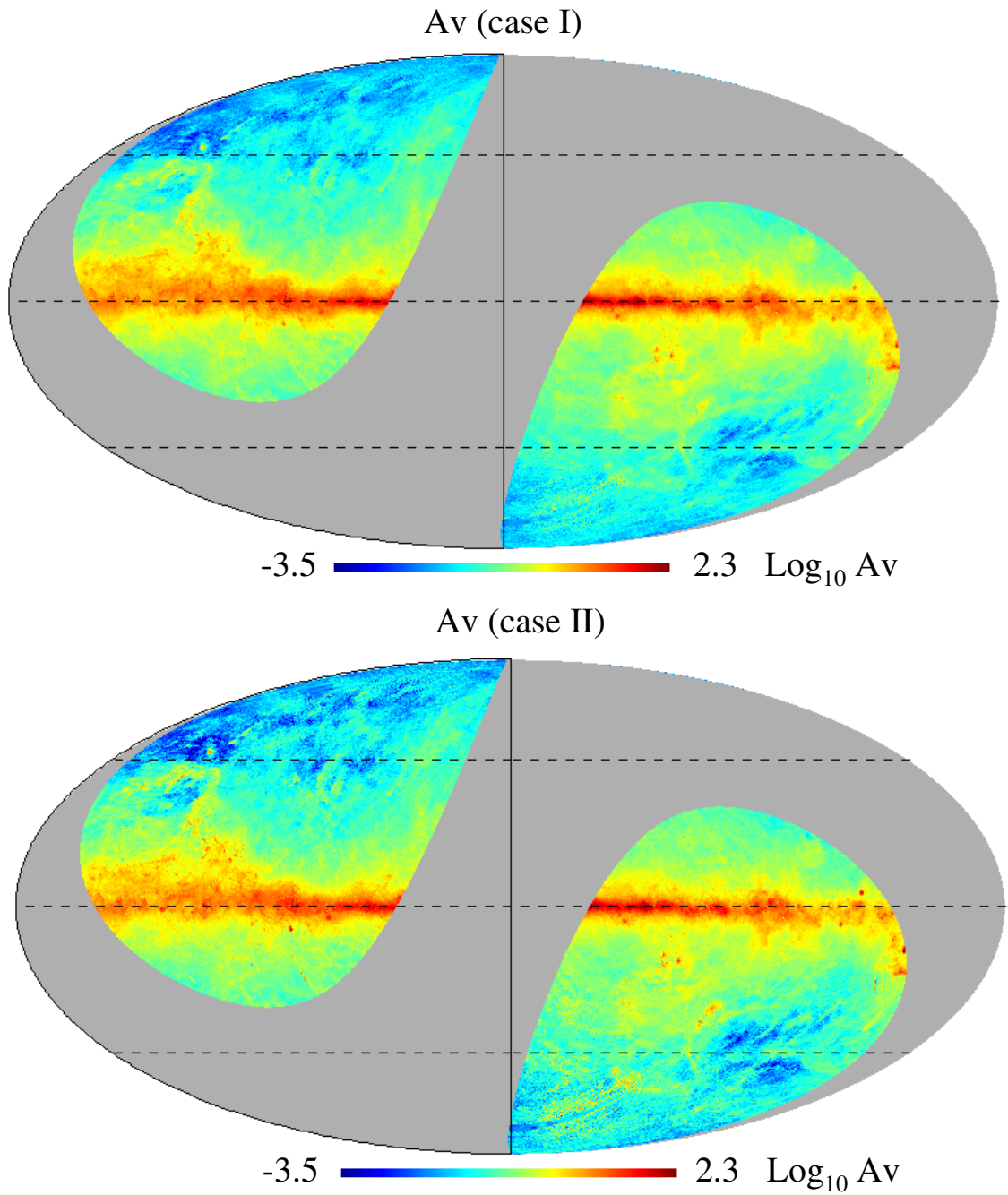}
  \end{center}
  \caption{Galactic extinction distributions at $V$ band [mag]. The extinction is shown in a common logarithm scale. Top and bottom columns show the results for cases I and II, respectively.}\label{fig:fig_5}
\end{figure}

\begin{figure}
  \begin{center}
    \FigureFile(90mm,90mm){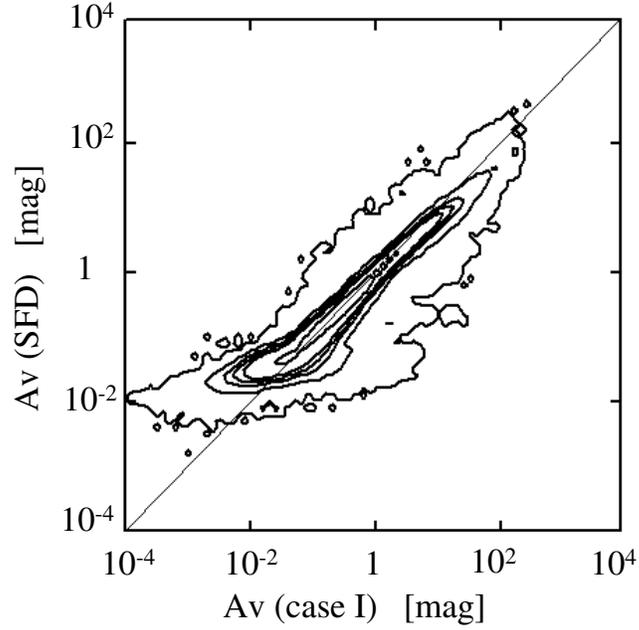}
  \end{center}
  \caption{Comparison between $A_{V}$ of this study and SFD. The systematic difference is 5\% and 21\%, for $A_{V}$ (SFD) $<$ 1 and $A_{V}$ (SFD) $>$ 1 mag, respectively. The systematic difference for $A_{V}$ (SFD) $>$ 1 mag may reflect the difference in the spatial resolution of the dust temperature. The wider dispersion for $A_{V}$ (SFD) $<$ 0.1 mag may reflect an SFD assumption (see text). The contour map includes the same parameters as those described in Figure 2. The bin size is set to 0.1 in logarithmic units. Contour lines are drawn according to the number of data points, including 0.5, 0.1N, 0.5N, N, 2N, and 10N in the diagram, where N = 1525 indicates the average number of data points. The solid line indicates a slope of unity. }\label{fig:fig_6}
\end{figure}

\begin{figure}
  \begin{center}
    \FigureFile(90mm,90mm){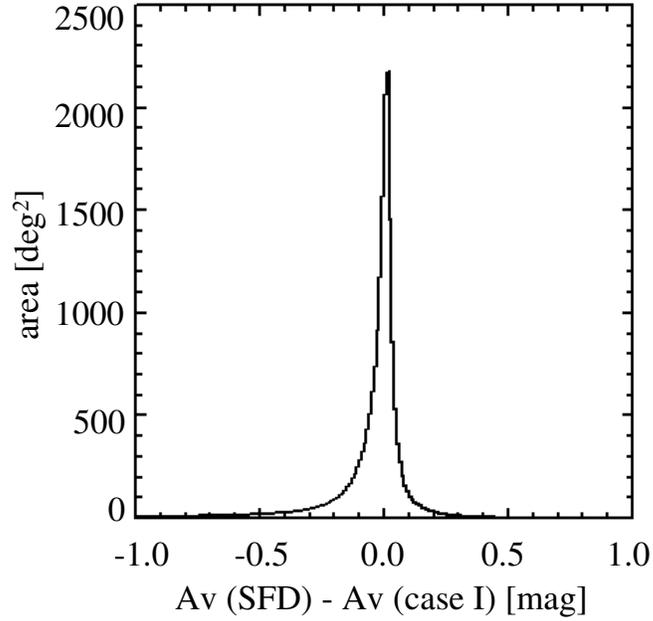}
  \end{center}
  \caption{Histogram of the $A_{V}$ difference between this study and SFD. Bin size is 0.01 mag. The mean, median, and standard deviation are $-0.28$, $-0.001$, and 2.29 mag, respectively.}\label{fig:fig_7}
\end{figure}

\begin{figure}
  \begin{center}
    \FigureFile(160mm,90mm){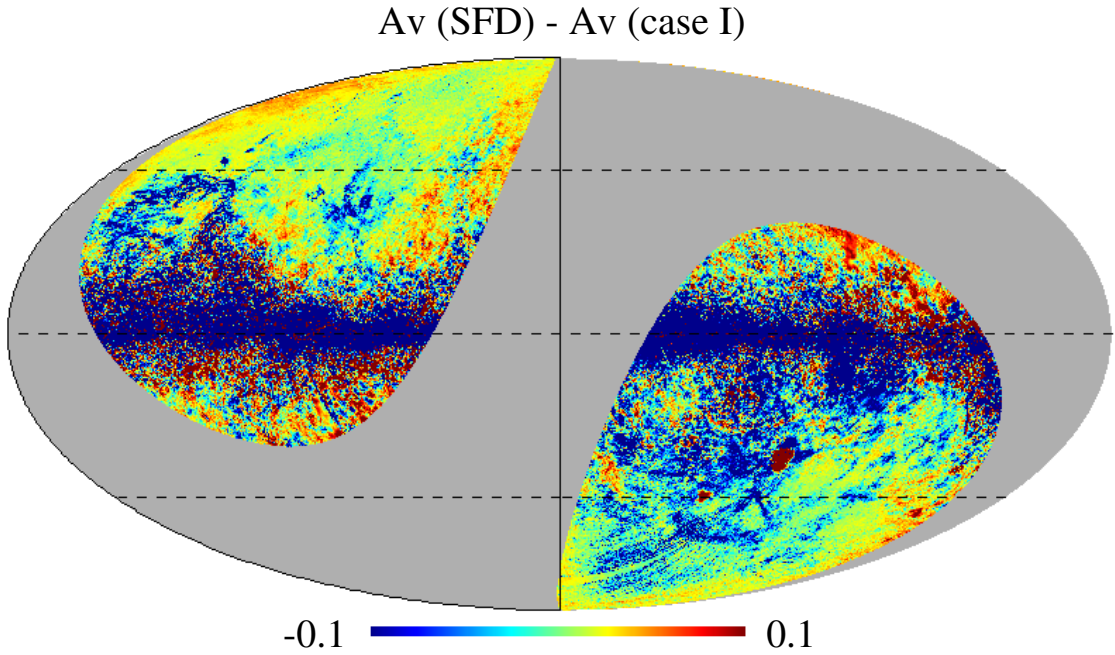}
  \end{center}
  \caption{Maps of the $A_{V}$ difference between this study and SFD. The range is $\pm$ 0.1 mag.}\label{fig:fig_8}
\end{figure}

\begin{figure}
  \begin{center}
    \FigureFile(160mm,90mm){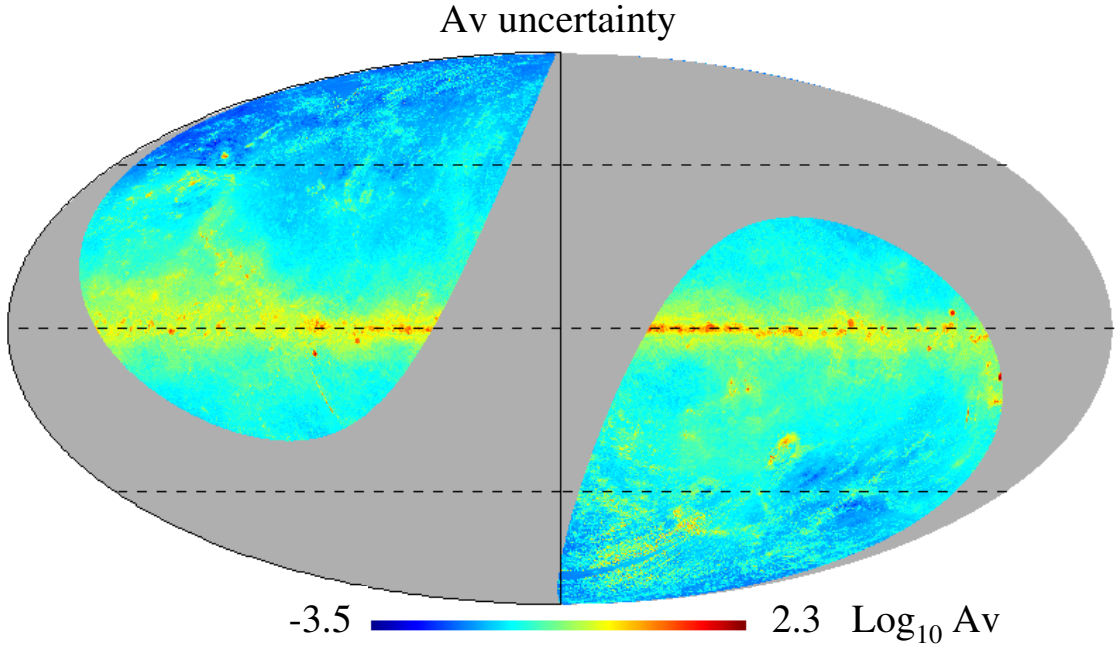}
  \end{center}
  \caption{Map of the $A_{V}$ uncertainty of our method estimated by Equation (5) [mag]. The value is shown in a common logarithm scale.}\label{fig:fig_9}
\end{figure}

\begin{figure}
  \begin{center}
    \FigureFile(160mm,90mm){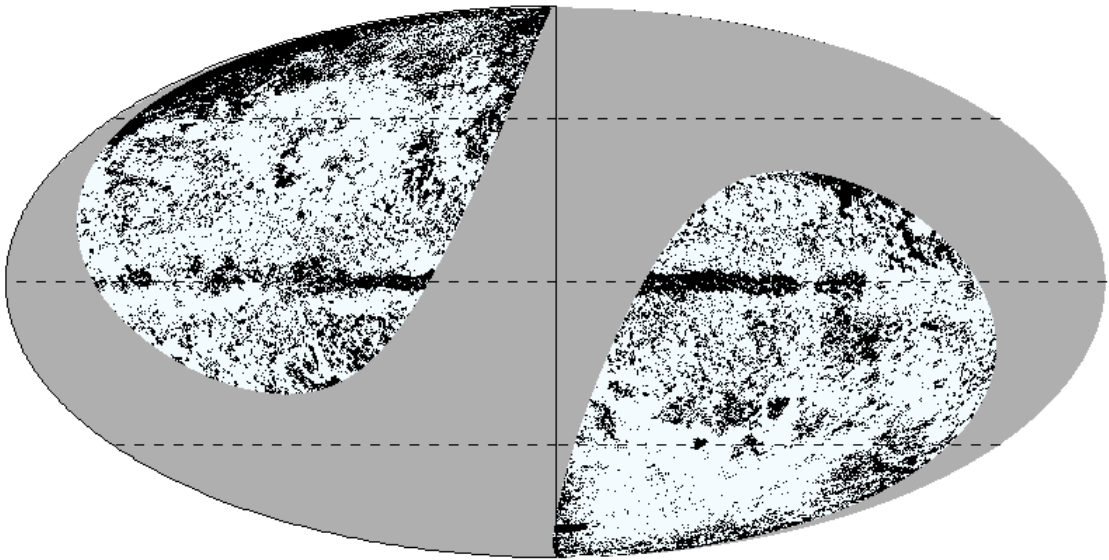}
  \end{center}
  \caption{Area in which the significant difference between this study and SFD is confirmed. Black shading represents areas in which the difference is larger than the $A_{V}$ uncertainty of our method. The area occupies 28\% of the region for $\mid \beta \mid \ > 30^\circ$.}\label{fig:fig_10}
\end{figure}

\begin{figure}
  \begin{center}
    \FigureFile(90mm,90mm){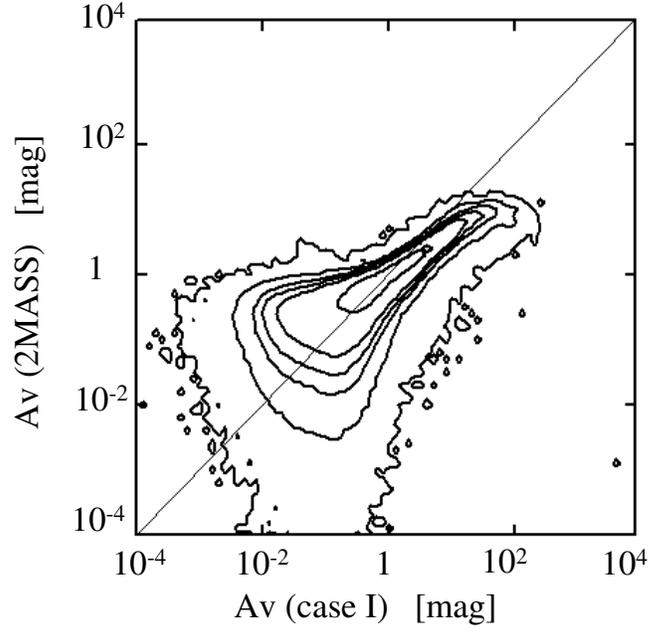}
  \end{center}
  \caption{Comparison of $A_{V}$ between this study and Dobashi (2011). The $A_{V}$ (2MASS) are saturated at $A_{V}$ (2MASS) $>$ 20 mag. The dispersion in the vertical direction at $A_{V}$ (case I) $<$ 1 mag may be caused by the low S/N of $A_{V}$ (2MASS) in high galactic latitudes. The solid line indicates a slope of unity. LMC and SMC are masked. The contour map includes the same parameters as those described in Figure 6. The average number of data points contained in each bin is 732.}\label{fig:fig_11}
\end{figure}

\begin{figure}
  \begin{center}
    \FigureFile(90mm,90mm){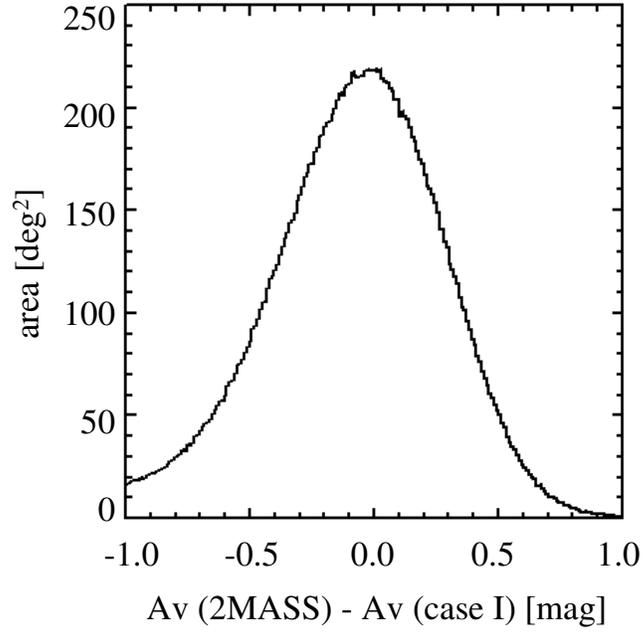}
  \end{center}
  \caption{Histogram of the $A_{V}$ difference between this study and Dobashi (2011). Bin size is 0.01 mag. The mean, median, and standard deviation of the $A_{V}$ difference are $-0.58$, $-0.09$, 3.69 mag, respectively. LMC and SMC are masked.}\label{fig:fig_12}
\end{figure}

\begin{figure}
  \begin{center}
    \FigureFile(160mm,90mm){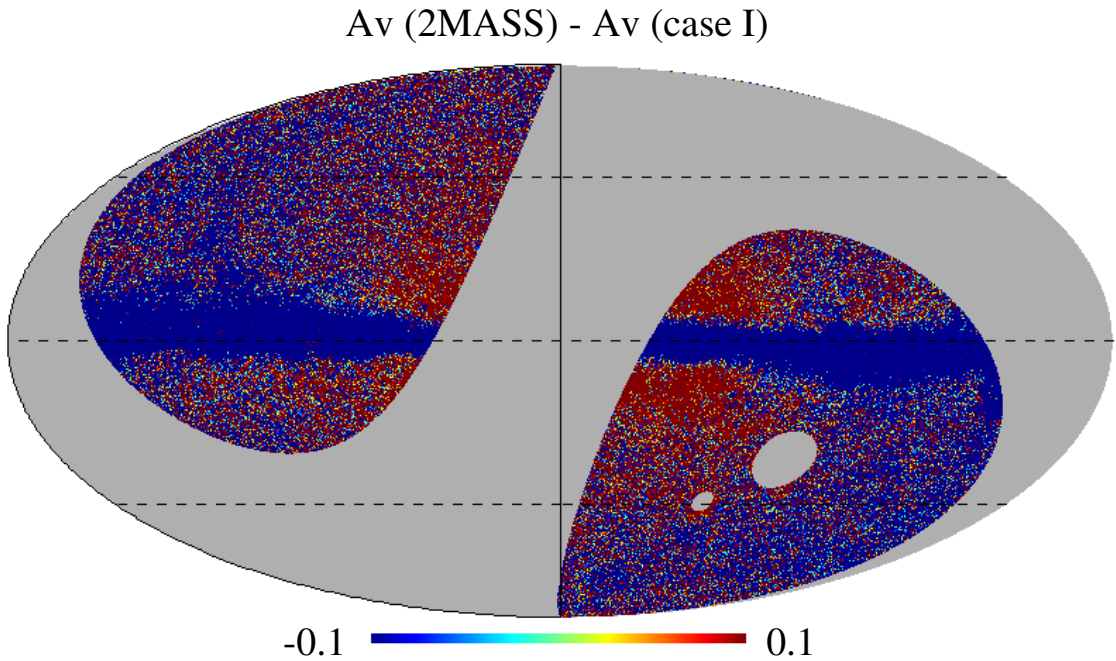}
  \end{center}
  \caption{Map of the $A_{V}$ difference between this study and Dobashi (2011). The range is $\pm$ 0.1 mag. LMC and SMC are masked.}\label{fig:fig_13}
\end{figure}

\clearpage

\begin{figure}
  \begin{center}
    \FigureFile(160mm,90mm){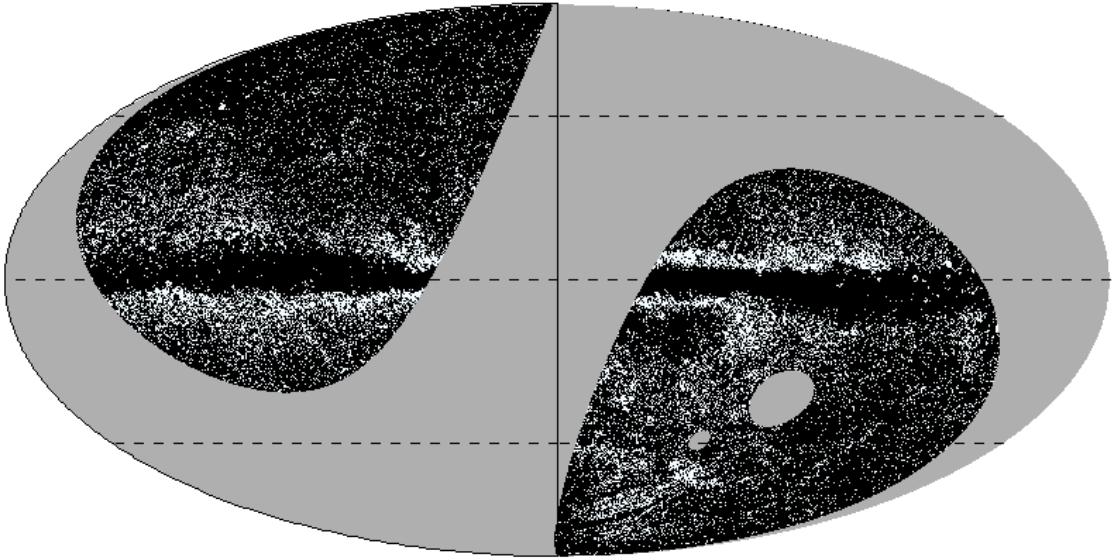}
  \end{center}
  \caption{Area in which the significant difference between this study and Dobashi (2011) is confirmed. Black shading represents areas in which the difference is larger than the $A_{V}$ uncertainty of our method. The area occupies 81\% of the region for $\mid \beta \mid \ > 30^\circ$. LMC and SMC are masked.}\label{fig:fig_14}
\end{figure}

\begin{table}[b]
\begin{center}
	\caption{Results of $A_{V}$ comparisons among this study, SFD, and Dobashi (2011).}\label{tab:tab_7}
	\begin{tabular}{ccccccc}
	\hline
	\hline
	& \shortstack{Effective \\ region} & Wavelength & \multicolumn{2}{c}{Spatial resolution} & \shortstack{1$\sigma$ $A_{V}$ difference \\ between this study } & \shortstack{Upper limit \\ for $A_{V}$}\\
	& & [$\mu$m] & \multicolumn{2}{c}{[arcmin]} & [mag] & [mag] \\
	\hline
	& & & $A_{V}$ & Dust temperature  &  & \\
	\hline
	This study & $\mid \beta \mid \ > 30^\circ$ & 100 \& 140 & 5 & 5 & -- & over 100 \\
	SFD &  All-sky & 100 \& 240 & 5 & 42 & 2.29 & over 100 \\
	Dobashi (2011) & All-sky & 1.25 \& 1.65 & 1 -- 12 & -- & 3.69 & 20 \\
	\hline
	\multicolumn{1}{@{}l}{\rlap{\parbox[t]{.95\textwidth}{\small
	The $A_{V}$ difference is compared with the result for case I; i.e., $A_{V}$ (case I).}}}
\end{tabular}
\end{center}
\end{table}

\end{document}